\documentclass[aps,onecolumn,12pt,secnumarabic,balancelastpage,amsmath,amssymb,nofootinbib]{revtex4}
\usepackage{graphicx}
\usepackage{dcolumn}
\usepackage{bm}
\usepackage{physics}
\usepackage{mathrsfs}
\usepackage{float}
\usepackage{soul, xcolor}
\usepackage{color}
\usepackage{hyperref}
\newcommand{\E}{\mathcal{E}}      
\newcommand{\D}{\mathscr{D}}      
\newcommand{\vect}[1]{\mathbf{#1}} 

\begin{document}

\preprint{APS/123-QED}

\title{Improving sensitivity to magnetic fields and electric dipole moments by using measurements of individual magnetic sublevels}

\author{Cheng Tang}
\author{Teng Zhang}%
\author{David S. Weiss}%
\affiliation{%
 Physics Department, The Pennsylvania State  University.\\
 104 Davey Laboratory, University Park, Pennsylvania 16802, USA
}%

\date{\today}
\setstcolor{red}

\begin{abstract}
We explore ways to use the ability to measure the populations of individual magnetic sublevels to improve the sensitivity of magnetic field measurements and measurements of atomic electric dipole moments (EDMs). When atoms are initialized in the $m=0$ magnetic sublevel, the shot-noise-limited uncertainty of these measurements is $1/\sqrt{2F(F+1)}$ smaller than that of a Larmor precession measurement. When the populations in the even (or odd) magnetic sublevels are combined, we show that these measurements are independent of the tensor Stark shift and the second order Zeeman shift. We discuss the complicating effect of a transverse magnetic field and show that when the ratio of the tensor Stark shift to the transverse magnetic field is sufficiently large, an EDM measurement with atoms initialized in the superposition of the stretched states can reach the optimal sensitivity. 
\end{abstract}


\maketitle


\graphicspath{{Figures/} }

\section{Introduction}

Precession of atomic angular momentum has been used to make magnetometers that approach the standard quantum limit \cite{Romalis, NMOR} and to search for permanent electric dipole moments (EDM) \cite{ACME, HgEDM, RaEDM, Commins}. Inherent sensitivity to angular momentum dependent quantities is optimal when a measurement is sensitive to the energy difference between the stretched states (the states in which $|m| = F$). For instance, preparation and observation of the time evolution of the superposition of stretched states can reach the optimal sensitivity. While it is relatively simple to prepare superpositions of stretched states in systems with $F=1/2$ or 1, it requires multi-photon processes to prepare the desired state in systems with $F>1$. One approach to building up the most sensitive state in atoms with $F>1$ is modulation of the optical pumping beam at $2F \omega_L$, where $\omega_L$ is the Larmor frequency, or its sub-harmonics \cite{Budker}. A multi-photon process is required during detection in this approach, which compromises its efficiency. It has been demonstrated that a quantum projection uncertainty near the optimum value can be obtained by precession from the $m=0$ magnetic sublevel, which can be prepared simply by optically pumping \cite{Heinzen}. If only the m=0 population is measured, the sensitivity depends strongly on the precession phase. Here we extend the work of \cite{Heinzen} to detection of all magnetic sublevels, using the technique of \cite{VLS}. We find three advantages compared to detection of the $m=0$ level alone. First, detection of all magnetic sublevels recovers the full sensitivity of the system, for all precession phases. Second, when the total population in the even magnetic sublevels is measured, it does not depend on quadratic energy shifts due to electric or magnetic fields that are perpendicular to the measurement axis. Third, we can construct a combination of the populations of magnetic sublevels to extract a pure sinusoidal signal.

In this paper, we first analyze individual-sublevel detection in a typical Larmor precession experiment with atoms initialized in one of the stretched states, and find that this does not offer better sensitivity than what is obtained by measuring the expectation value $\expval{F_x}$. We then analyze measurements of individual magnetic sublevels in precession from the $m=0$ state and show how measurements of individual magnetic sublevels improve the sensitivity. We next discuss the effect of transverse magnetic fields on these precession measurements. Finally, we discuss the application of these ideas to EDM measurements.

\section{Individual Magnetic sublevels of Larmor precession}

\begin{figure}[H]\centering
	\includegraphics[width=3in]{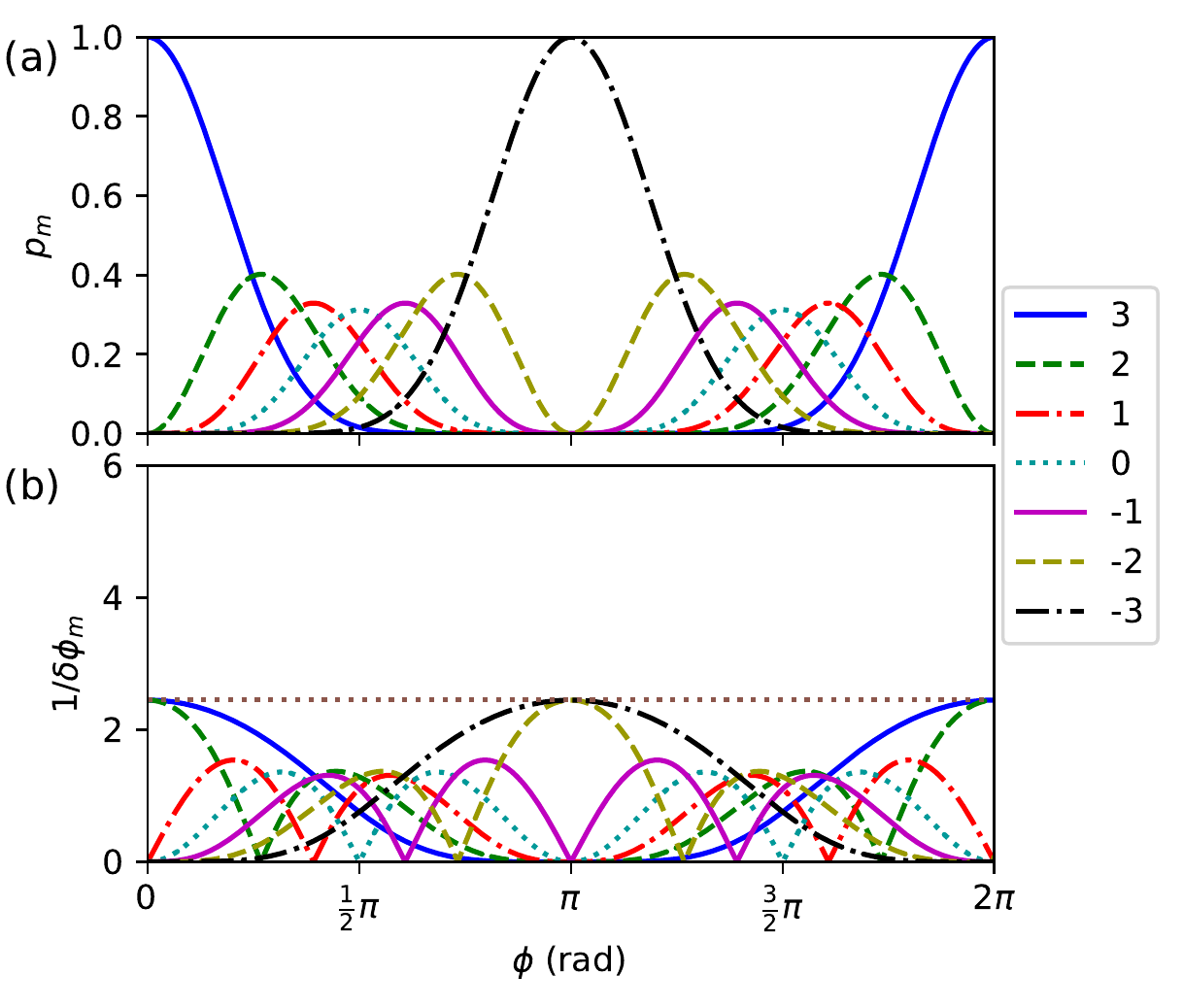}
    \caption{(Color online). Precession from the state $\ket{3, 3}$. \textbf{(a)} Probabilities to be in individual magnetic sublevels $m$ (color coded) as a function of phase. \textbf{(b)} The inverse of the uncertainties obtained with individual magnetic sublevels as a function of phase. The dotted horizontal line is the inverse of the uncertainty obtained by measuring the projection $\expval{F_x}$. The upper bound of this figure ($1/\delta\phi_m=6$) is the inverse of the uncertainty of the optimal measurement.}
    \label{fig:Precession33}
\end{figure}

A typical Larmor precession measurement detects the expectation value of some component of the  angular momentum $\expval{\vect{F}}$. It has maximum contrast when atoms are initialized in a stretched state.  We consider precession from a stretched state, say $\ket{\psi}=\ket{F, m=F}_x$, for an atom with angular momentum $F>1$. The expectation value of the angular momentum precesses in a magnetic field, $B_z$, at the Larmor frequency, $\omega_L=g_F\mu_BB/\hbar$, where $g_F$ is the Land\'e $g_F$-factor and $\mu_B$ is a Bohr magneton. $\omega_L$ is proportional to the separation between adjacent magnetic sublevels, which for large $F$ is much smaller than the largest energy difference in the problem, which is the separation between stretched states. The shot-noise-limited phase or frequency uncertainty obtained by detecting $\expval{F_x}$ scales as $1/\sqrt{2F}$. The question naturally arises whether one can improve the sensitivity of a Larmor precession measurement by using the ability to measure the populations of individual sublevels. The probabilities of detecting atoms in individual magnetic levels are $p_m=\bra{\psi'}\ket{m}_x\bra{m}_x\ket{\psi'}$, where $\ket{\psi'}=e^{-iHt/\hbar}\ket{\psi}$ is the state after precession and $\ket{m}_x$ are the eigenstates of $F_x$. To evaluate $p_m$, we write the initial state $\ket{\psi}$ in the $\vect{z}$ basis, in which the Hamiltonian $H=\hbar\omega F_z$ is diagonal. Specifically, the $p_m$ are given by:
\begin{equation}\label{core}
p_m=\left |\sum_{m'm''}d^\dag_{mm'}(\pi/2)d_{m'm''}(\pi/2)\bra{m''}_x\ket{\psi}e^{-im'\phi}\right |^2,
\end{equation}
where $d_{m'm''}(\pi/2)$ are the Wigner rotation matrix elements, and $\phi=\omega \tau$ is the phase accumulated over the precession time of $\tau$.
We will use $F=3$ as an example, where the initial state $\ket{\psi}$ is $\ket{3,3}_x$. The probabilities of detecting each magnetic sublevel after precession are plotted in Fig.~\ref{fig:Precession33}a. The phase uncertainty obtained by these detections are given by the quantum projection noise divided by the slope of the signal with respect to the phase:
\begin{equation}
	\delta\phi_m=\sqrt{\expval{P_m^2}-\expval{P_m}^2}/\left |\frac{dp_m}{d\phi}\right|,
	\label{eq:uncertainty}
\end{equation}
where $P_m=\ket{m}\bra{m}$ are the projection operators for the individual magnetic sublevels. The inverse of phase uncertainties, $1/\delta\phi_m$, are plotted in Fig.~\ref{fig:Precession33}b. For comparison, we also calculate the phase uncertainty obtained by measuring $\expval{F_x}=\sum_m p_m m$. The phase uncertainty obtained by detecting $\expval{F_x}$ is:
\begin{equation}
	 \delta\phi_{\expval{F_x}}=\sqrt{\expval{F_x^2}-\expval{F_x}^2}/\left|\frac{d\expval{F_x}}{d\phi}\right|=1/\sqrt{2F},
\end{equation}
as indicated by the dotted horizontal line in Fig.~\ref{fig:Precession33}b. It is evident from Fig.~\ref{fig:Precession33}b that the sensitivity obtained by detecting any single magnetic sublevel is never any better than the sensitivity obtained by detecting $\expval{F_x}$.

The sensitivities obtained by detecting individual sublevels can be combined to give a net sensitivity better than that obtained from any single magnetic sublevel. In the absence of correlations, sensitivities from independent measurements, $\delta\phi_i$, can be combined as follows,
\begin{equation}
    \frac{1}{\delta\phi_c^2}=\sum_i \frac{1}{\delta\phi_i^2},\label{eq:combine}
\end{equation}
where $\delta\phi_c$ is the combined uncertainty. However, the uncertainties $\delta\phi_m$ as given by Eq.~(\ref{eq:uncertainty}) and plotted in Fig~\ref{fig:Precession33} are correlated with each other. We can take into account the correlations by analyzing the measurement of all the magnetic sublevels as a sequence of measurements, keeping track of all outcomes based on their probabilities. In those cases where the atom is in the measured sublevel, the measurement is complete. When the atom is not in that sublevel, the probability is shared among the remaining sublevels, and the next measurement is independent of the prior ones. The sensitivities of a sequence of measurements can then be combined using Eq.~(\ref{eq:combine}). Of course, one expects the same result regardless of the order in which this imagined sequence of measurements is made.

We calculate the full sensitivity obtained by detecting all magnetic sublevels by calculating the result for a sequence of $2F+1$ measurements as follows. Consider a normalized state after precession of phase $\phi$:
\begin{equation}
    \ket{\psi}=\sum_{m=-F}^{F} a_m(\phi) \ket{m},
\end{equation}
where $a_m(\phi)$ are the amplitudes of the state $\ket{m}$. Suppose we detect the atom in $\ket{m=F}$ as the first in the sequence of measurements. The phase uncertainty of the first measurement follows directly from Eq. (\ref{eq:uncertainty}) and can be written as:
\begin{equation}
    {\delta \phi_1}=\frac{\sqrt{p_{m=F}(\phi)-p_{m=F}(\phi)^2}}{|\dot{p}_{m=F}|},
\end{equation}
where $p_{m=F}(\phi)=|a_{m=F}(\phi)|^2$ is the probability to be in $m=F$ and $\dot{p}_{m=F}={d p_{m=F}(\phi)}/{d\phi}$ is the slope. The remaining probability, $1/(1-p_{m=F}(\phi))$, is in the undetected states. The new state is:
\begin{align}
    \ket{\psi'}=\frac{1}{\sqrt{1-p_{m=F}(\phi)}}\sum_{m=-F}^{F-1} a_m(\phi) &\ket{m}\\
    \equiv \sum_{m=-F}^{F-1} a'_m(\phi) &\ket{m},
\end{align}
where the prefactor normalizes the collapsed state and the primed amplitudes $a'_m(\phi)=a_m(\phi)/{\sqrt{1-p_{m=F}(\phi)}}$ are the amplitudes after normalization. Then we carry out the next measurement, say on $\ket{m=F-1}$. The phase uncertainty obtained by measuring $\ket{m=F-1}$ from the collapsed state is:
\begin{equation}
    \delta\phi_2=\frac{1}{\sqrt{1-p_{m=F}(\phi)}}\frac{\sqrt{p'_{F-1}(\phi)-p'_{F-1}(\phi)^2}}{|\dot{p'}_{F-1}|},
\end{equation}
where ${\sqrt{1-p_{m=F}(\phi)}}$ accounts for the probability of any of these outcomes occuring, $p'_{F-1}=|a_{F-1}(\phi)|^2/({1-p_{F}(\phi)})$ is the probability of measuring $m=F-1$ from the collapsed state $\ket{\psi'}$ and ${\dot{p'}_{F-1}}={d p'_{m=F-1}(\phi)}/{d\phi}$ is the slope. In this expression for $\delta\phi_2$, the previous measurement has altered both the slope and the projection uncertainty. We repeat the above procedure until all the remaining probability is in a single magnetic sublevel, which will at last be detected with 100\% probability, and therefore offers no measurement sensitivity. The sensitivities of the sequence of measurements can be combined using Eq. (\ref{eq:combine}).
\begin{figure}[htb]\centering
	\includegraphics[width=3in]{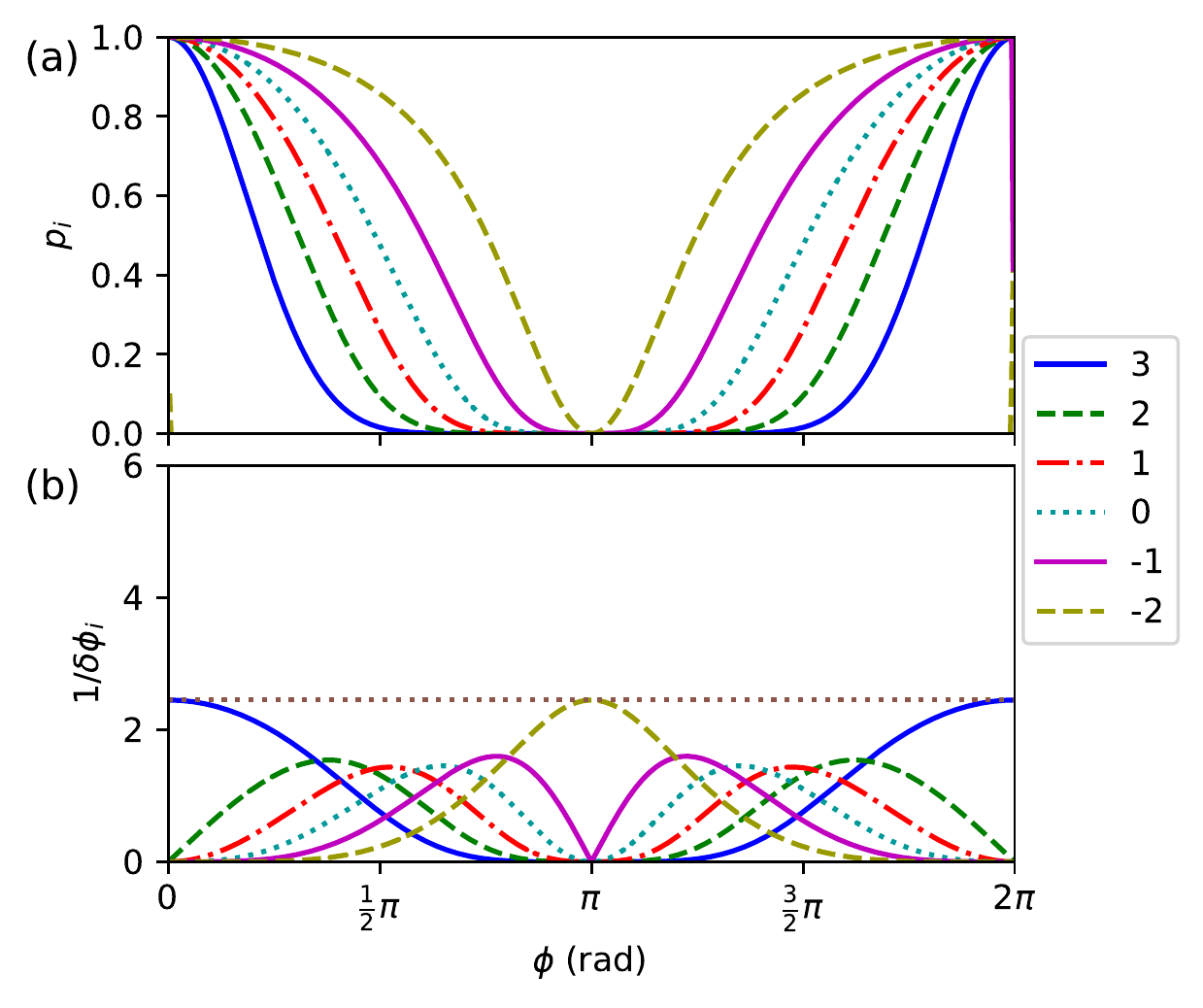}
    \caption{(Color online). Sequential analysis of individual sublevel measurements after precession from the state $\ket{3, 3}$. We measure the magnetic levels in a decreasing sequence of $m$ starting from $m=+3$. \textbf{(a)} The probabilities of finding the atom in a given magnetic sublevel after it was not found in previous measurements. \textbf{(b)} The corresponding inverse phase uncertainties. Not shown in this figure is the last measurement, which detects $m=-3$ with 100\% probability and no sensitivity. The dotted horizontal line is the inverse of the combined uncertainty of the independent measurements of all magnetic sublevels.}
	\label{fig:sequential detection}
\end{figure}

We apply the above method to analyzing the full sensitivity obtainable in precession from $\ket{3,3}$. Suppose we measure the magnetic levels in a decreasing sequence of $m$ starting from $m=+3$. The probability of finding the atoms in a given magnetic level after previous null measurements and the inverse of the associated phase uncertainty are plotted in Fig~\ref{fig:sequential detection}. Compared with results in Fig~\ref{fig:Precession33}, the state collapses alter all measurements except the first. The combined uncertainty of these independent measurements is exactly $1/\sqrt{2F}$, identical to the uncertainty obtained by measuring the expectation value $\expval{F_x}$. Detection of the populations in individual magnetic sublevels does not improve sensitivity in this case. But, as we show in the next section, such measurements are needed to take full advantage of precession from an $m=0$ state.

\section{Individual magnetic sublevels precessed from the $m=0$ state}

\begin{figure}[htb]\centering
	\includegraphics[width=3in]{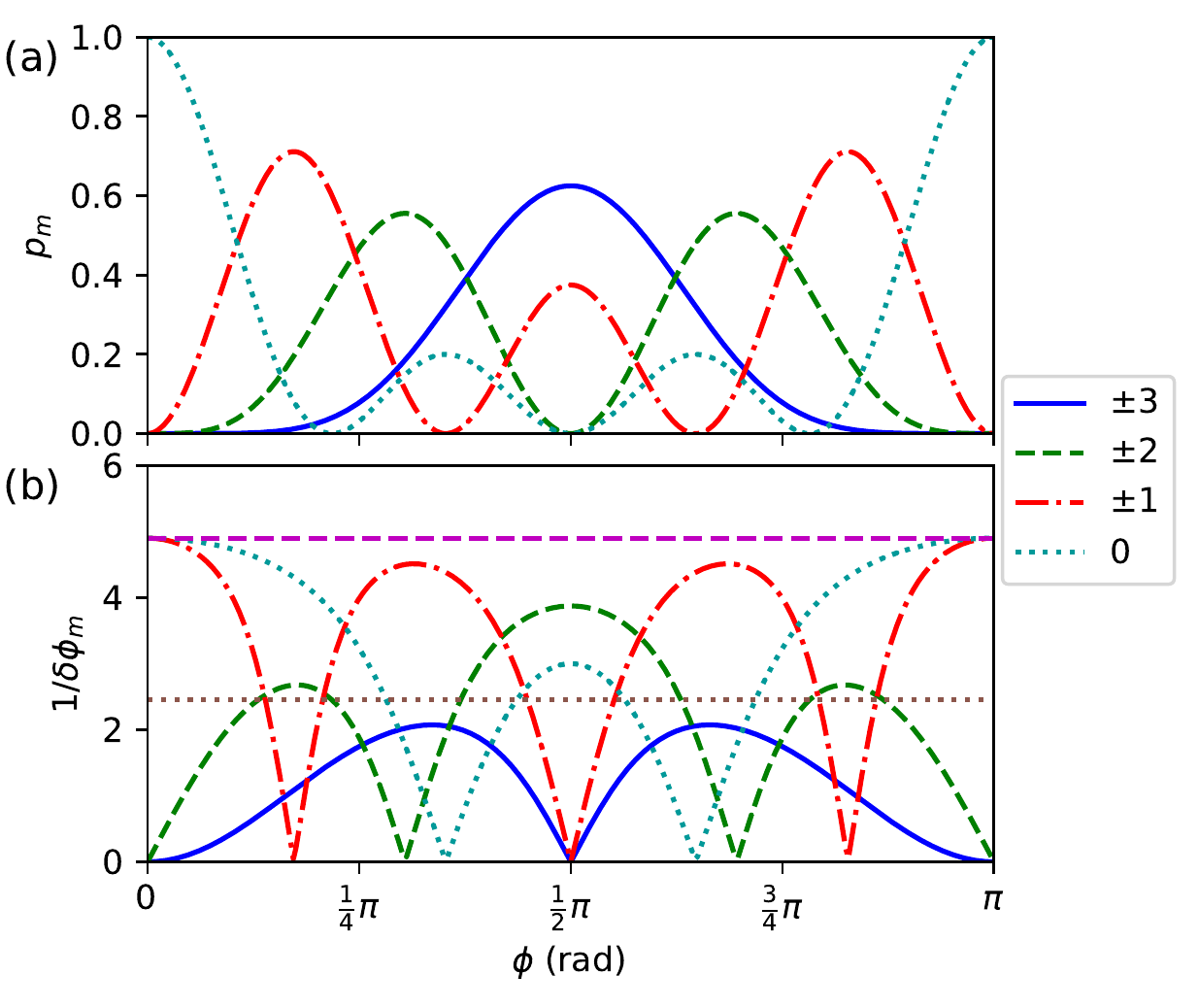}
    \caption{(Color online). Precession from the state $\ket{3, 0}$. Both sub-figures are periodic with respect to the phase with a minimum period of $\pi$. \textbf{(a)} The probabilities to be in the different magnetic sublevels $m$ as a function of phase. Because the probability to be in $m$ is identical to the probability to be in $-m$, the probabilities to be in $m$ and $-m$ are summed in this figure. \textbf{(b)} The inverse of the uncertainties obtained by detecting individual magnetic sublevels as a function of phase. The dashed horizontal line is the inverse of the combined uncertainty obtained by independent measurements of all magnetic sublevels. The dotted horizontal line is the inverse of the combined uncertainty obtained in precession from a stretched state.}
    \label{fig:Precession30}
\end{figure}

When atoms are initialized in the $m=0$ level, the expectation value $\expval{\vect{F}}$ is zero and remains zero throughout the precession, so there can be no conventional Larmor precession measurement. Of course, the individual magnetic sublevels do evolve. The phase or frequency sensitivity obtained by measuring $\ket{F, 0}$ with atoms initialized in the $\ket{F, 0}$ state was demonstrated by Xu and Heinzen \cite{Heinzen}. Here we consider how the populations of all magnetic sublevels in the $F=3$ hyperfine level evolve after being initialized in $m=0$. The precession of individual state populations and the inverse of the associated phase uncertainties are shown in Fig.~\ref{fig:Precession30}.

The smallest phase uncertainty obtained by detecting the probability to be in the state $\ket{3,0}$ is half of what is obtained by Larmor precession and 22.5\% larger than what is obtained with an optimal measurement. A similar result for $F=4$ was first shown in \cite{Heinzen}. This sensitivity is available only around a phase of 0 or $\pi$, where the slope $dp_0/d\phi$ is close to 0. To make a precision measurement of the magnetic field or the EDM, it is often desirable to scan the phase (either by scanning the bias magnetic field or the precession time) over a larger range. This goal can be realized by combining measurements of all magnetic sublevels, as illustrated in the previous section.  The full sensitivity obtained by independent measurements of all the magnetic sublevels is equal to the best sensitivity of an $m=0$ measurement, but at all phases. In the following two subsections, we introduce two ways of combining individual magnetic sublevel measurements to yield a single  potentially useful fringe.

\subsection{Probability to be in Even Magnetic sublevels}

\begin{figure}[H]\centering
	\includegraphics[width=3in]{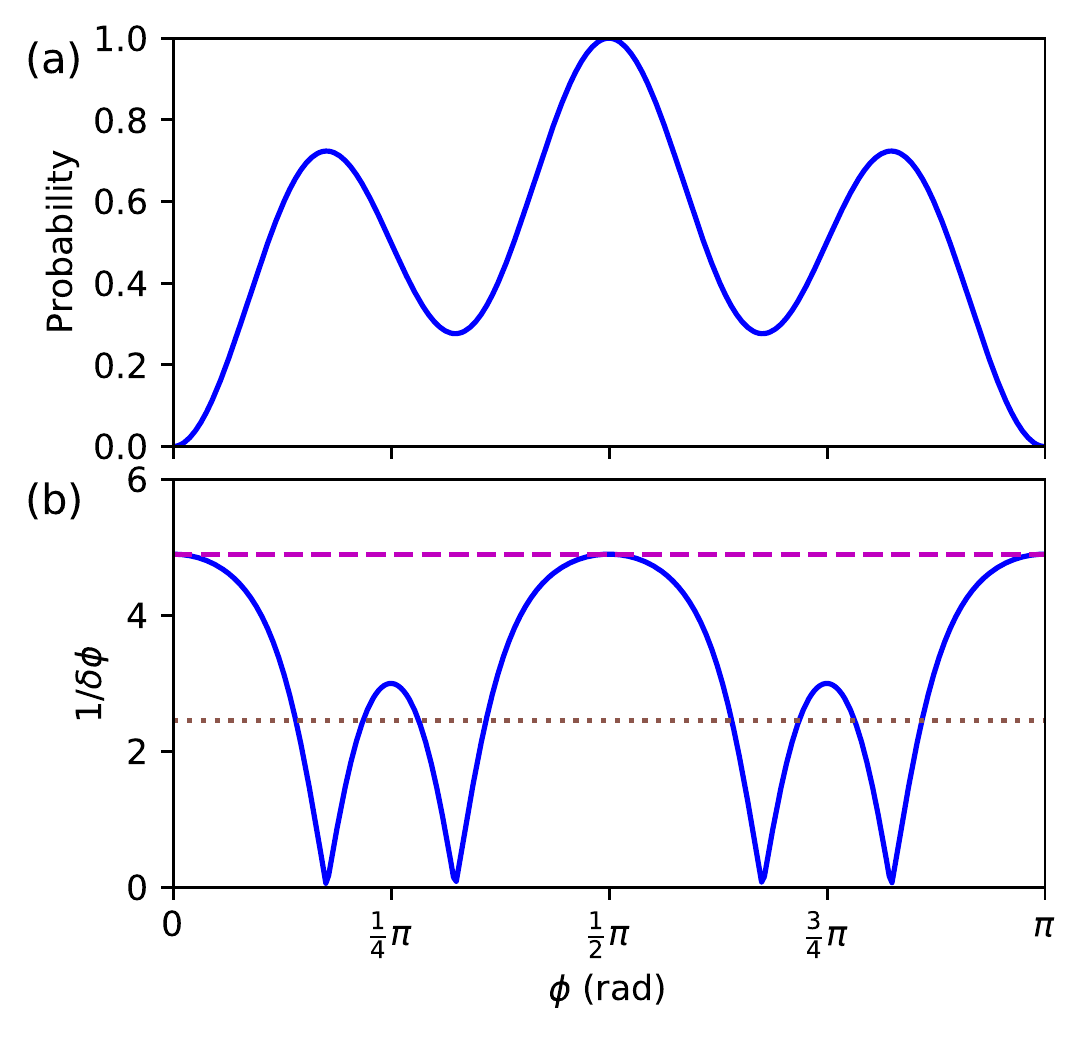}
    \caption{ \textbf{(a)} The sum of the probabilities in the even magnetic sublevels during precession from the state $\ket{3, 0}$. \textbf{(b)} The inverse of the uncertainty from this measurement (solid line). The best sensitivity of the even magnetic sublevels is the same as the best sensitivity of the $m=0$ state shown in Fig.~\ref{fig:Precession30}. The dashed horizontal line is the inverse of the combined uncertainty obtained by independent measurements of all magnetic sublevels. The dotted horizontal line is the inverse of the uncertainty obtained in typical Larmor precession.}
	\label{fig:even}
\end{figure}

The sensitivity obtained by taking the sum of the probabilities to be in the even magnetic sublevels is shown in Fig.~\ref{fig:even}. The best sensitivity is comparable to the result for  $\ket{3,0}$. Using this combination of individual measurements seems simpler than keeping separate track of the individual sublevel evolutions, and though the fringe shape is not as simple as a sinusoid, it is not very complicated, containing just two Fourier components. More importantly, these fringes are unaffected by the tensor Stark shift, the second order Zeeman shift and any interaction that is even with respect to $m$ in the orthogonal direction, even though the evolution of any individual magnetic sublevel does depend on these shifts. Proof of this insensitivity is given in Appendix~\ref{app:even}. The insensitivity to quadratic shifts requires that the measurement axis to be orthogonal to the fields that give rise to the quadratic energy shift.
Of course, since the probabilities to be in even and odd magnetic sublevels sum up to 1, we could as well have used the odd magnetic sublevels for this discussion.

\begin{figure}[!htb]\centering
	\includegraphics[height=.7\textheight]{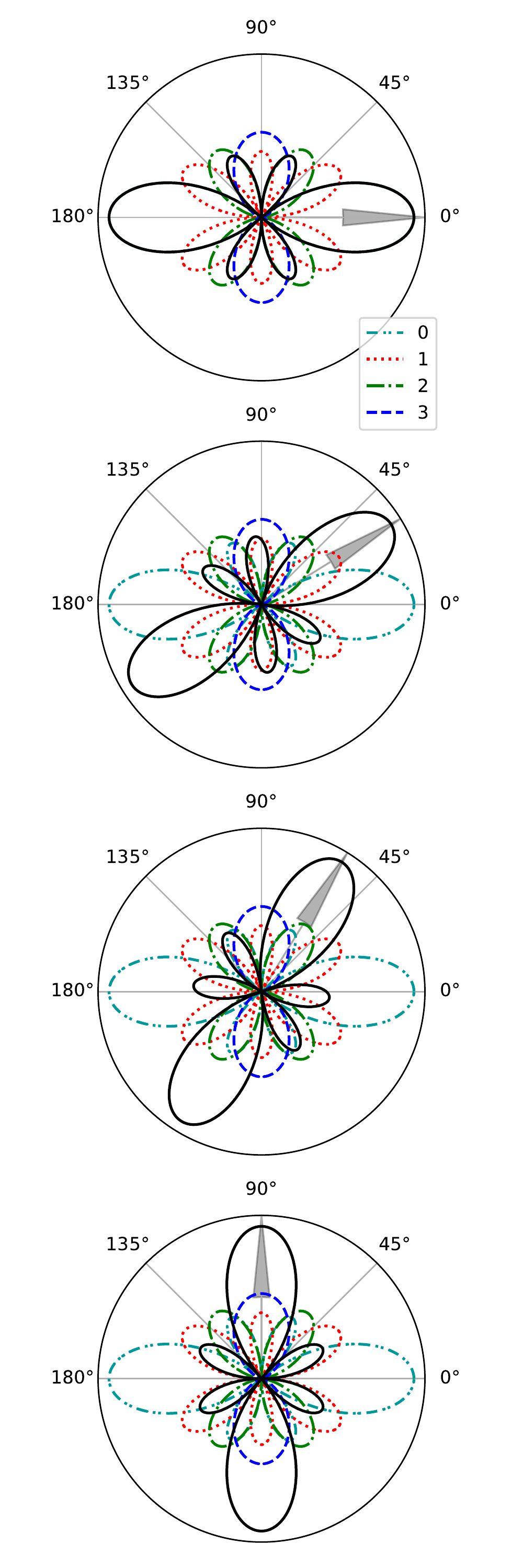}
    \caption{(Color online). The precession of spherical harmonics. The spherical harmonic of the state that precesses starting from $\ket{3, 0}_x$ is shown by the solid black line and its symmetry axis is represented by the gray arrow. The spherical harmonics of the eigenstates of $F_x$ are shown with dashed or dotted lines. Each sub-figure shows the unprecessed states $\ket{3,m}_x$ along with a snapshot of the precessing state at a phase corresponding to an extremum in Fig.~\ref{fig:even}. The extrema correspond to points of relatively good overlap with successive $|m|$ values.}
	\label{fig:spherical harmonics}
\end{figure}

In order to get a physical sense for why keeping track of the even (or odd) populations gives heightened sensitivity, we can visualize the precession of the $m=0$ state of a general atomic angular momentum, $F$, by considering the spherical harmonics associated with an orbital angular momentum of the same value. In Fig.~\ref{fig:spherical harmonics}, we show the precession of an initial $\ket{3, 0}_x$ state at the precession phases corresponding to the first 4 extrema in Fig.~\ref{fig:even}, and compare them to the other spherical harmonics in the x-basis. As the state precesses, the plane of maximum amplitude in the rotating state overlaps in turn with the lobes of the various basis states. Of course, the axis of symmetry of the precessing state rotates, so the precessing state never quite looks like the comparison states. Still, the results of the rigorous calculation presented in Fig.~\ref{fig:even} are made graphically clear in these pictures. The precessed state has considerable overlap in turn with the $\ket{3, \pm1}$ states, the $\ket{3,  \pm2}$ states, and then the $\ket{3,  \pm3}$ states, accounting for the high frequency component of the fringe in Fig.~\ref{fig:even}.

\subsection{Single harmonic from linear combination of magnetic sublevels}
Another notable combination of individual magnetic sublevel evolution is the one that yields the $2F$\textsuperscript{th} order polarization moment. The probability curves in the various $m$ levels are generally composed of the sum of the 0\textsuperscript{th} to the $2F$\textsuperscript{th} harmonics of the Larmor frequency. We can single out the highest order harmonic using linear combination of the probabilities to be in various $m$ levels with different weights, $\alpha_m$. In the case of $F=3$, the measurement operator that yields the hexacontatetrapole is
\begin{equation}
P_{2F}=\sum_m\alpha_m \ket{m}\bra{m}=\ket{0}\bra{0}-\frac{8}{7}(\ket{1}\bra{1}+\ket{-1}\bra{-1})+\frac{11}{7}(\ket{2}\bra{2}+\ket{-2}\bra{-2})
\end{equation}

\begin{figure}[H]\centering
	\includegraphics[width=3in]{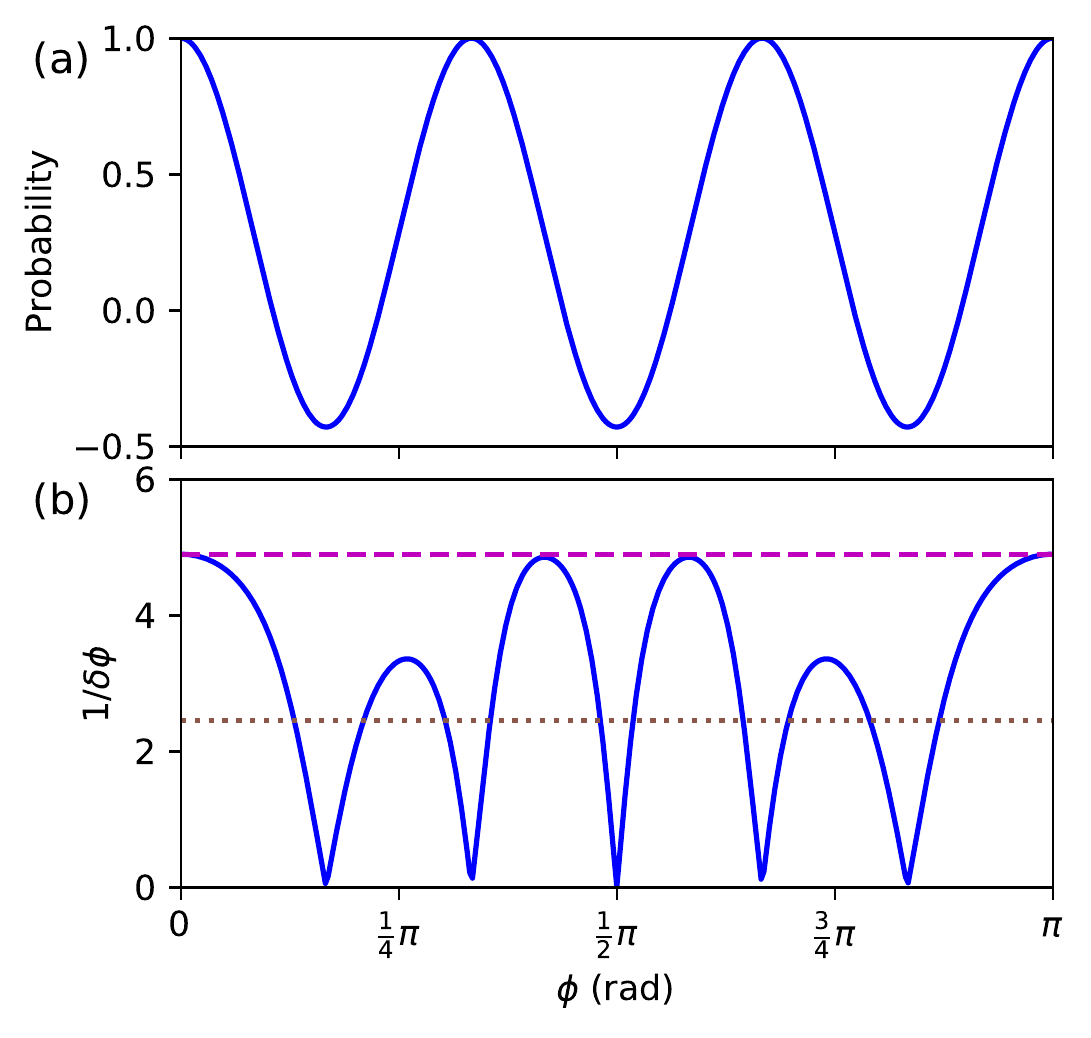}
    \caption{\textbf{(a)} Precession of the 6th harmonic starting from the state $\ket{3, 0}$. \textbf{(b)} The inverse of the uncertainty from this measurement (solid line). The dashed horizontal line is the inverse of the combined uncertainty obtained by independent measurements of all magnetic sublevels. The dotted horizontal line is the inverse of the uncertainty obtained in typical Larmor precession.}
\end{figure}

\subsection{Extension to higher integer and half integer $F$}
\begin{figure}[H]\centering
	\includegraphics[width=3in]{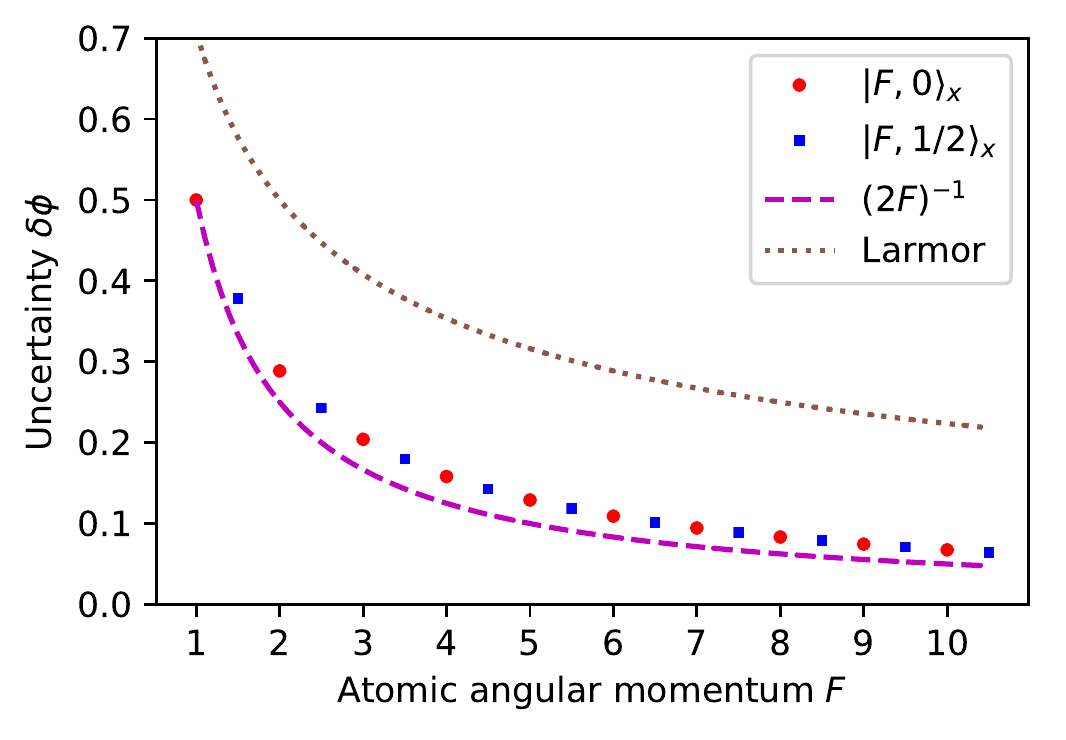}
    \caption{(Color online). Shot-noise-limited phase uncertainty scaling with $F$. The circles are the combined uncertainties obtained by independent measurements of all magnetic sublevels in precession from the state $\ket{F, 0}$ for integer $F$, and the squares are the combined uncertainties in precession from $\ket{F, 1/2}$ for half integer $F$. The dashed line is the uncertainty obtained with an optimal measurement, and the dotted line is the uncertainty obtained in typical Larmor precession. }\label{higher integer F}
\end{figure}

The above results can be extended to higher integer angular momentums in a straightforward manner. When precessing from the $\ket{F, 0}$ state, the shot-noise-limited phase uncertainty scales as {$1/\sqrt{2F(F+1)}$}. This is {$1/\sqrt{F+1}$} times what is obtained in typical Larmor precession starting from a stretched state and $\sqrt{2F/(F+1)}$ times what is obtained with an optimal measurement, as illustrated in Fig.~\ref{higher integer F}.

The reason that precession from the state  $\ket{F, 0}_x$ is more sensitive than precession from $\ket{F, F}_x$ lies in the fact that the state $\ket{F, 0}_x$ in the $\vect{z}$ basis has the strongest amplitudes in the stretched states among all eigenstates of $F_x$. The optimal measurement involves creating a superposition of the $m=\pm F$ levels, $(\ket{+F}+\ket{-F})/\sqrt{2}$, for which the shot-noise-limited uncertainty scales as $(2F)^{-1}$. We will discuss preparation of this superposition below. In general, it is less straightforward than preparing the eigenstate $\ket{F, 0}_x$. As can be seen from Fig.~\ref{higher integer F}, there is not much inherent sensitivity loss associated with using the more simply prepared state.

\begin{figure}[htb]\centering
	\includegraphics[width=3in]{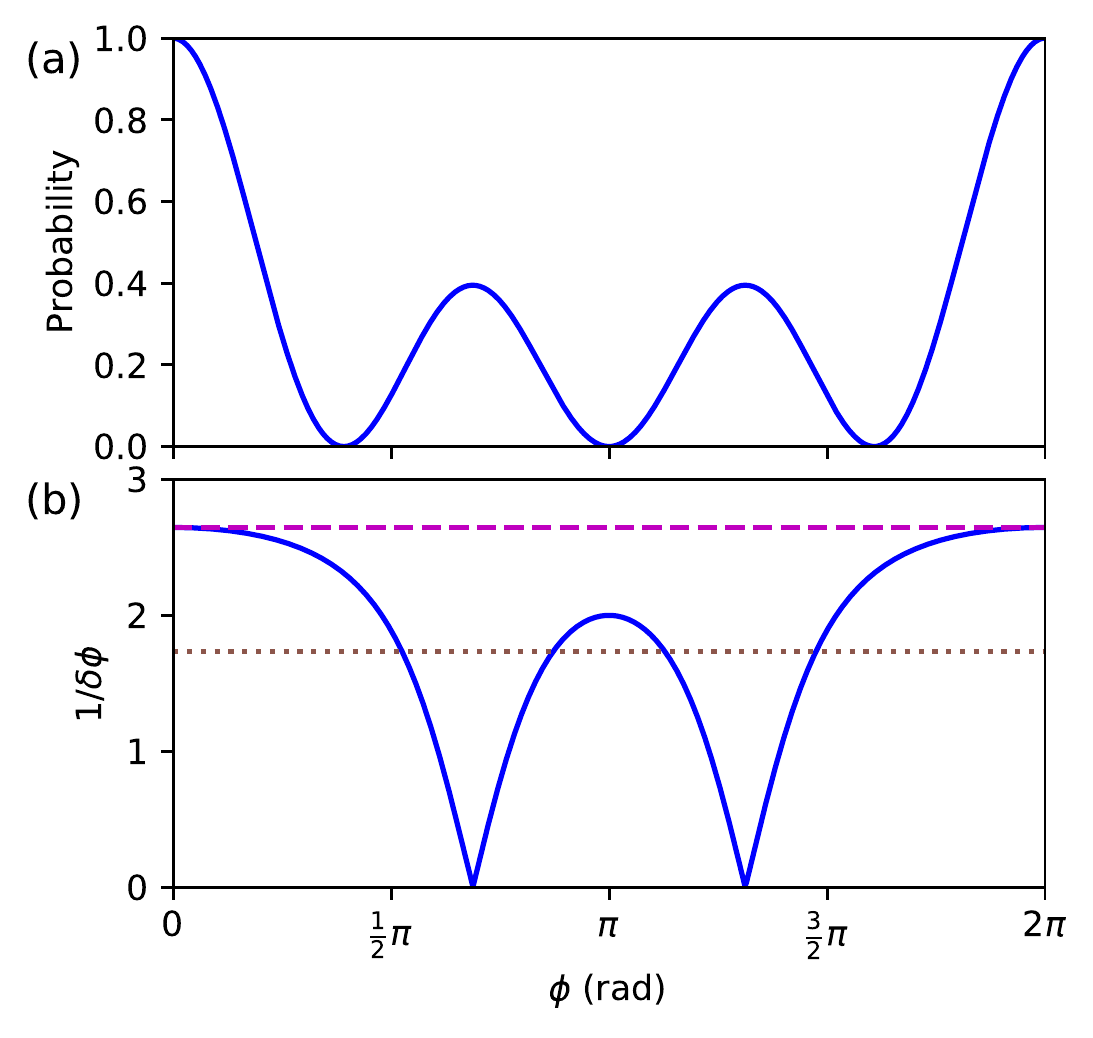}
    \caption{\textbf{(a)} The probability of measuring the state $\ket{3/2, 1/2}_x$ in precession from $\ket{3/2, 1/2}_x$. \textbf{(b)} The inverse of the uncertainty from this measurement (solid line). The dashed horizontal line is the inverse of the combined uncertainty obtained by independent measurements of all magnetic sublevels. The dotted horizontal line is the inverse of the uncertainty obtained by measuring the expectation value $\expval{F_x}$ in precession from a stretched state.}
    \label{fig:F3_2m0}
\end{figure}

To extend these calculations to half integer angular momentums we prepare atoms in $\ket{F, 1/2}$ (or equivalently $\ket{F, -1/2}$). The precession of the state $\ket{F, 1/2}$ is somewhat qualitatively similar to that of $\ket{2F, 0}$, as illustrated by the precession of $\ket{3/2, 1/2}$ shown in Fig.~\ref{fig:F3_2m0} compared to the $\ket{3, 0}$ curve in Fig.~\ref{fig:Precession30}. The best shot-noise-limited uncertainty obtained by using the the state $\ket{F, 1/2}$ scales as $1/\sqrt{2F(F+1-1/(4F))}$. Interestingly, these sensitivities neatly interleave the results for $\ket{2F, 0}$ evolution in integer $F$ systems (see Fig.~\ref{higher integer F}).

Atoms with large magnetic moments are inherently sensitive to magnetic fields, but Larmor precession does not take full advantage. For example, the shot-noise-limited uncertainty from Larmor precession of Dy, where $F=21/2$ in the ground state, is 4.58 times what can be obtained with an optimal measurement. If Dy is prepared in the $m=1/2$ state instead, the smallest uncertainty obtained by measuring $m=1/2$ alone or by measuring the evolution of all sublevels, is only a factor of 1.35 away from the optimal sensitivity.

\section{Effect of a transverse field on a measurement of magnetic field}

\begin{figure}[H]\centering
	\includegraphics[width=6in]{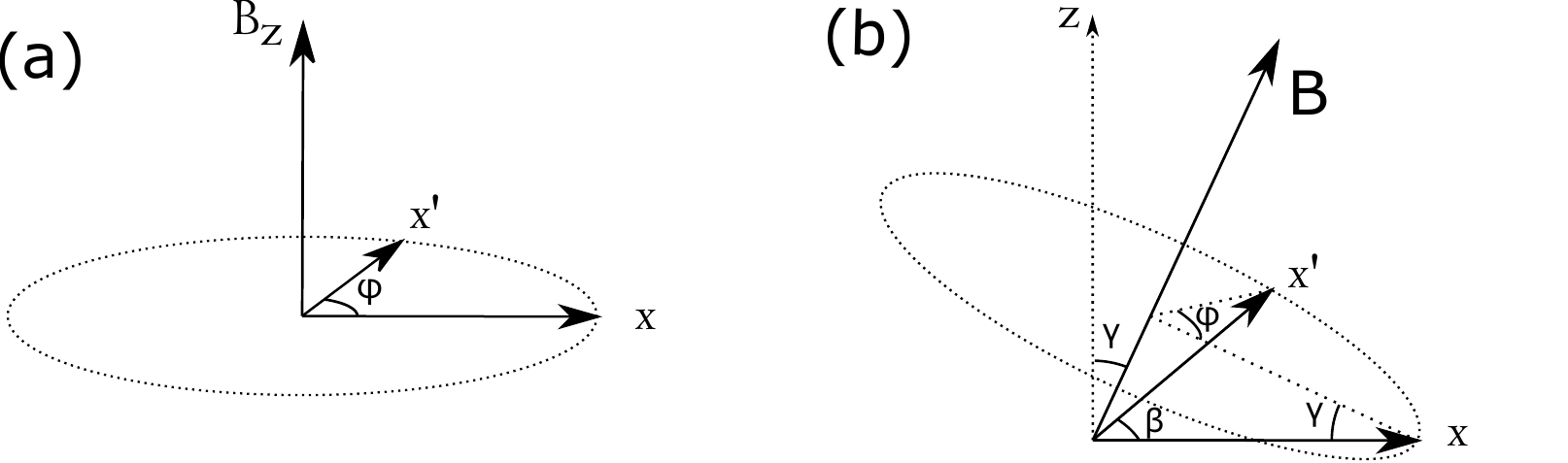}
    \caption{Schematic of angular momentum precession. \textbf{(a)}Angular momentum initialized in one of the eigenstates of $F_{x}$ precesses in a magnetic field along $\vect{z}$. \textbf{(b)} Angular momentum initialized in one of the eigenstates of $F_{x}$ precesses  in a magnetic field at an angle $\gamma$ from the $\vect{z}$ axis . A magnetic level in the $\vect{x}$ basis evolves into the corresponding level in the $\vect{x'}$ basis, where $\vect{x'}$ is related to $\vect{x}$ by a rotation of $\phi$ around $\vect{B}$.}
    \label{vector}
\end{figure}

The discussion so far has been limited to the ideal scenario (Fig.~\ref{vector}a) where atoms are initialized in one of the magnetic levels in the $\vect{x}$ basis and the magnetic field $\vect{B}$ is along a direction perpendicular to $\vect{x}$, say $\vect{z}$. In general, the magnetic field will not be perfectly aligned with the $\vect{z}$ axis but at angle $\gamma$ from the $\vect{z}$ axis as shown in Fig.~\ref{vector}b. The transverse component of the magnetic field is given by $B_\perp=Bsin(\gamma)$. A magnetic level in the $\vect{x}$ basis evolves into the corresponding level in the $\vect{x'}$ basis, where $\vect{x'}$ is related to $\vect{x}$ by a rotation of $\phi$ around $\vect{B}$. The measurement on the evolved state in the original basis of $\vect{x}$ is solely determined by the angle $\beta$ between $\vect{x}$ and $\vect{x'}$. The angle $\beta$ differs from the precession phase $\phi$ when $\gamma$ is nonzero, and is related to $\phi$ through the following equation:
\begin{equation}
sin(\beta/2)=sin(\phi/2) cos(\gamma).
\end{equation}
Although $\phi$ ranges from 0 to $2\pi$, $\beta$ does not have values between $\pi-2\gamma$ and $\pi+2\gamma$.

\begin{figure}[htb]\centering
	\includegraphics[width=3in]{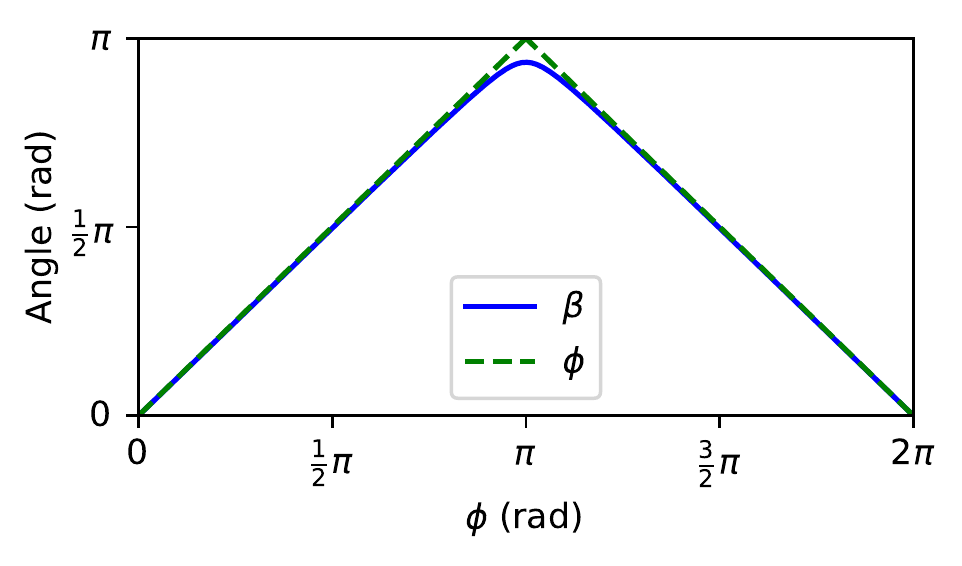}
    \caption{(Color online). $\beta$ versus $\phi$ for $sin(\gamma)=0.1$.}
    \label{fig:BetaPhi}
\end{figure}

\begin{figure}[htb]\centering
	\includegraphics[width=3in]{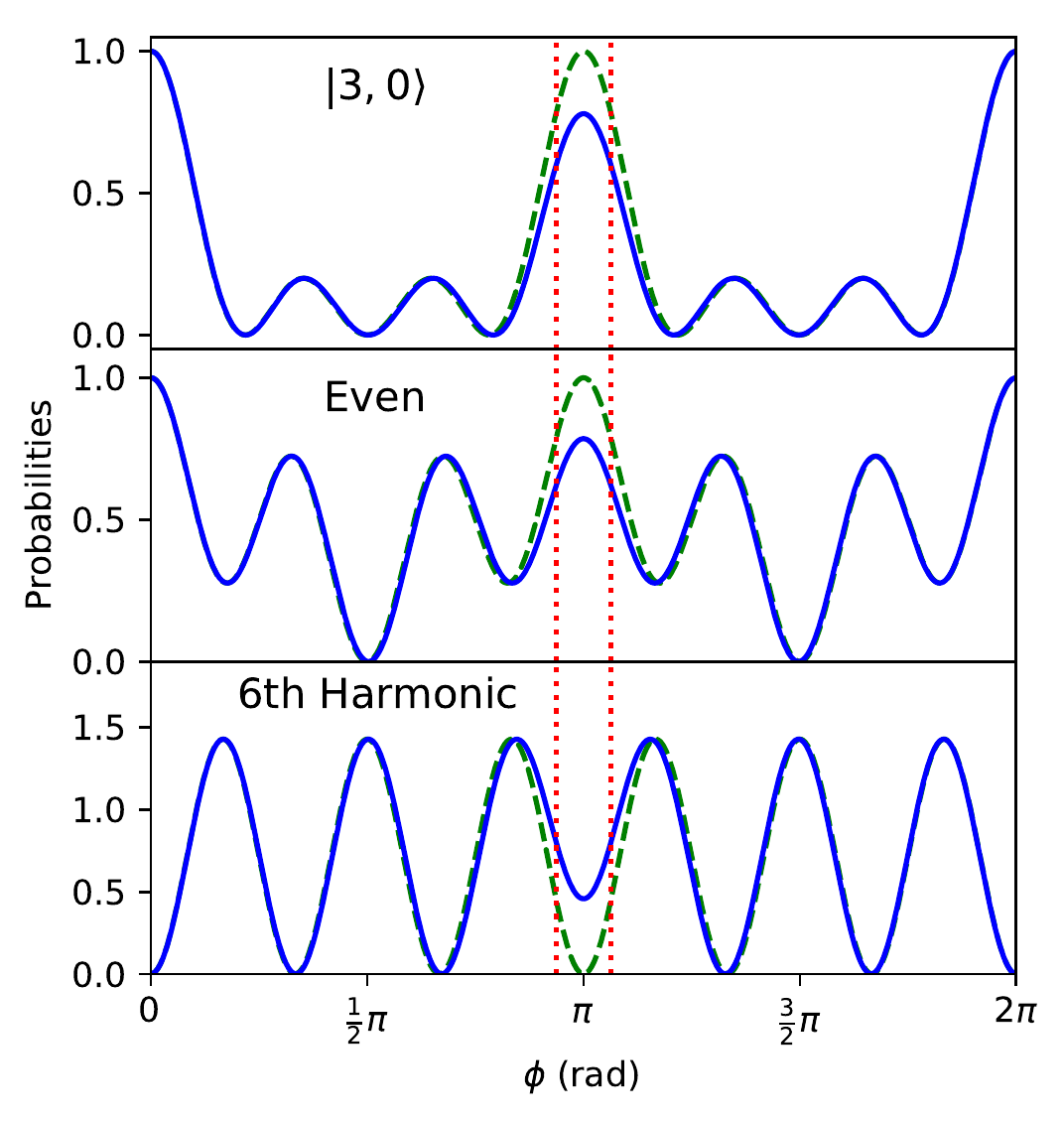}
    \caption{(Color online). Precession in the presence of a transverse magnetic field assuming $sin(\gamma)=0.1$ (solid line) and in the absence of a transverse magnetic field (dashed line). The vertical dotted lines mark $\pi-2\gamma$ and $\pi+2\gamma$.}
    \label{fig:transverseB}
\end{figure}

To illustrate how this effects fringe shapes, we will compare precession with $sin(\gamma)=0.1$ to precession without $B_{\perp}$. Fig.~\ref{fig:BetaPhi} shows the relationship between $\beta$ and $\phi$ for $sin(\gamma)=0.1$. The two angle variables are essentially equivalent except near the narrow range that $\beta$ does not reach. In Fig.~\ref{fig:transverseB} we plot, for both $sin(\gamma)=0.1$ and $B_{\perp}=0$, the evolution from the state $\ket{3, 0}_x$ in the $\vect{x}$ basis of the $m=0$ state, the even magnetic sublevels and the 6th harmonic. The two evolutions are nearly the same except in the narrow range where $\beta$ diverges from $\phi$.

\section{Effect of a transverse field in an EDM measurement}\label{sec:EDM}

For the purposes of an EDM measurement, one approach to preparing the superposition of stretched states that can give an optimal measurement is to exploit the tensor structure of the electric-field-induced tensor Stark shift. With a large enough electric field and small enough magnetic field, the same magnitude $m$ levels are nearly degenerate, while the degeneracy among $|m|$ pairs is lifted. The stretched state superposition can be produced by coherently driving atoms from the $m=0$ state using an oscillating transverse magnetic field that contains frequency components corresponding to all the $\Delta m=1$ transitions. For example,
the tensor Stark shift for the $F=3$ hyperfine level of cesium is given by \cite{Tensor2006, Tensor2007}:
\begin{equation}
{E}_\E(m)=\frac{1}{2}\alpha_2\frac{3m^2}{28}\E^2,
\end{equation}
where $\E$ is the electric field $\alpha_2$ is the tensor polarizability and the zero of the energy has been shifted such that ${E}(m=0)=0$. One can efficiently drive the atoms from $m=0$ to the superposition state of $(\ket{3,+3}+\ket{3,-3})/\sqrt{2}$  using three frequency components, corresponding to the transitions $m=0$ to $\pm1$, $\pm1$ to $\pm2$ and $\pm2$ to $\pm3$ \cite{Zhu}.

\begin{figure}[h]\centering
	\includegraphics[width=3in]{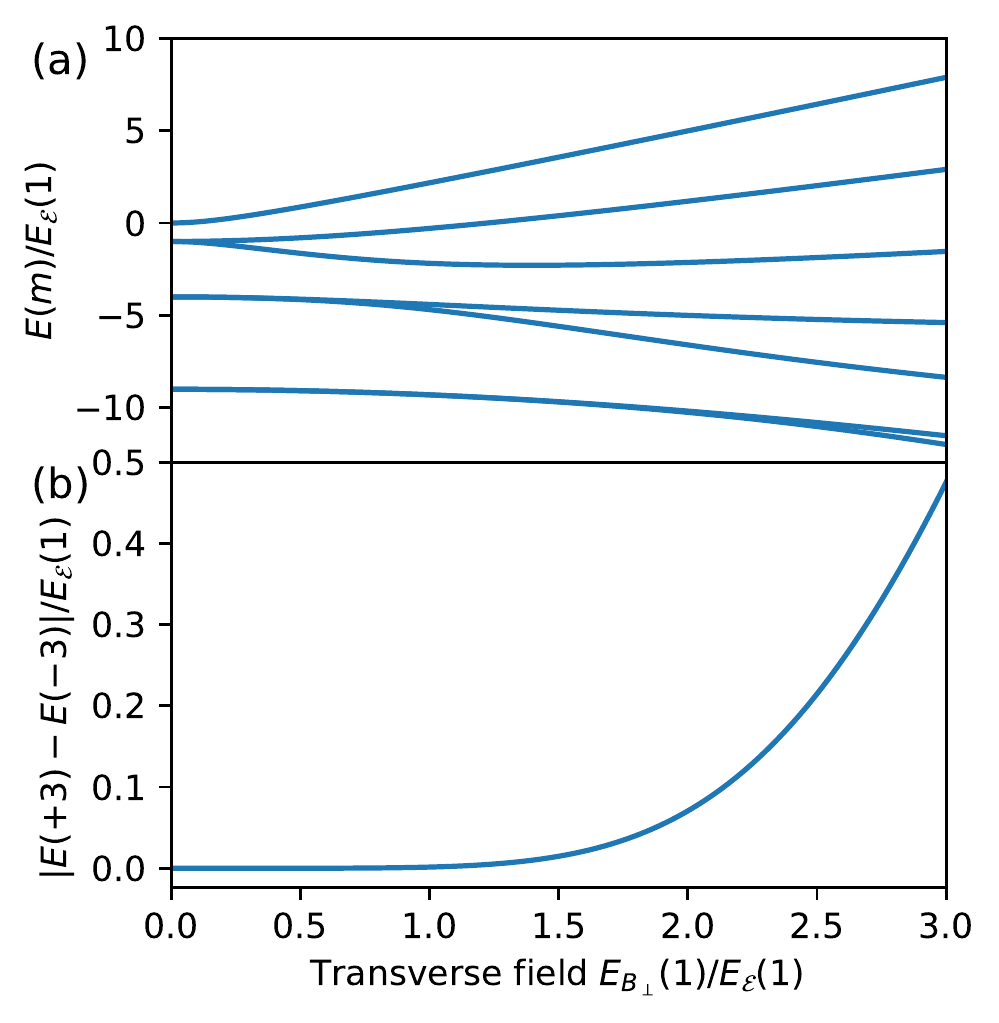}
    \caption{(a) Eigenenergies scaled to the tensor Stark shift as a function of transverse magnetic field in unit of linear Zeeman shift scaled to the tensor Stark shift, for $F=3$ (b) Energy difference between the stretched states as a function of the transverse field. Figure adapted from \cite{Zhu}}
    \label{fig:eigenval}
\end{figure}

Any transverse magnetic field, ${E}_{B_\perp}$, or fictitious field due to vector light shifts\cite{Zhu}, deforms the energy level structure, as different magnetic sublevels in the original basis become coupled. The deformation affects the stretched states least, since lifting their degeneracy involves $2F$ magnetic dipole couplings. The larger $F$ is, the more resistant the degeneracy is to being lifted, as illustrated in Fig.~\ref{fig:eigenval}a for $F=3$. For ${E}_{B_\perp}(1)<<{E}_\E(1)$, the energy difference between the stretched states varies as the sixth power of $B$ (see Fig.~\ref{fig:eigenval}b). The degeneracy of the stretched states, and thus the ease with which the superposition of stretched states can be prepared, is compromised once ${E}_{B_\perp}(1)$ is on the order of ${E}_\E(1)$. In contrast, it is easy to prepare the $m=0$ state, so if transverse magnetic fields cannot be well enough controlled, it may be advantageous to accept the modest loss in precision described earlier in this paper, and measure the EDM with a precession measurement from the $m=0$ state.

\begin{figure}[H]\centering
	\includegraphics[width=3in]{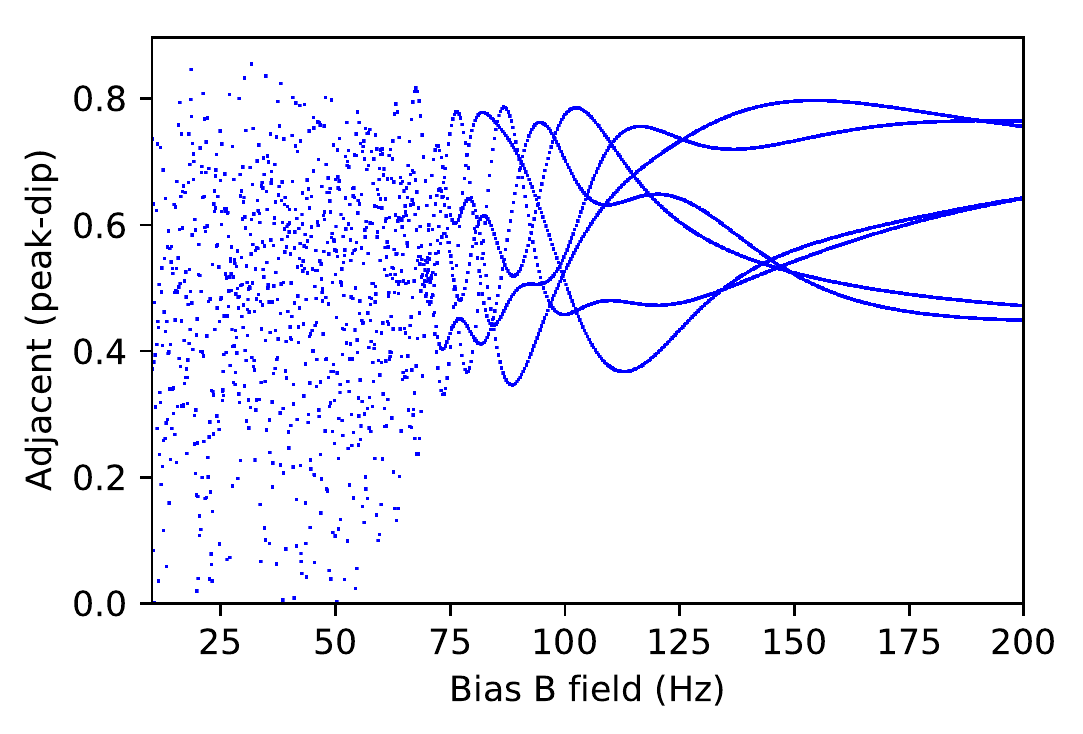}
    \caption{Differences between adjacent peaks and valleys in a scan of the bias magnetic field. We assume a tensor Stark shift of ${E}_\E(1)=5$ Hz, a transverse magnetic field of 10 Hz and a precession time of 3 seconds.}
    \label{fig:EDM with tranverse B}
\end{figure}

When $B_\perp=0$, using the precession of the $m=0$ state to measure the EDM still requires decoupling the tensor Stark shift from the linear shift of the magnetic sublevels. This can be done using one of two strategies:
\begin{enumerate}
  \item  $\E$ and $\tau$ can be chosen so that the phase accumulated due to the tensor Stark shift is an integer multiple of $2\pi$. Then the system behaves as if there was no tensor Stark shift at all. 
  \item The sum of the probabilities to be in the even (or the odd) magnetic sublevels can be measured, since it is independent of the tensor Stark shift in the orthogonal direction (see Appendix~\ref{app:even}).
\end{enumerate}
Both of these strategies are compromised when $B_\perp\neq0$.  In contrast, a bias field $B_z$ adds an offset to the precession (and might add noise), but does not otherwise change the precession signal. It is therefore possible to increase $B_z$ to mitigate the detrimental effect of $B_\perp$. To investigate the minimum bias field required to maintain useful signal fringes, we have carried out a numerical study for the concrete scenario where the tensor Stark shift is ${E}_\E(1)=5$ Hz and the transverse field is 10 Hz. Taking partial advantage of the above 2 strategies, we set the precession time to be 3 seconds and measure the probability to be in the even magnetic sublevels. Fig.~\ref{fig:EDM with tranverse B} shows the differences between the adjacent peaks and valleys as we scan the bias field from 10 Hz to 200 Hz. Below 60 Hz, the points are erratic, illustrating that there is no stable fringe pattern in that region. Above 60 Hz, the points start to form lines. These lines converge to 3 values around a bias field of ~200 Hz. These 3 values correspond to the differences between adjacent peaks and valleys of the stable probability curve in Fig.~\ref{fig:even}. It takes a bias field of about 20 times the transverse field for there to be a robust precession signal.

\section{Conclusion}
In this paper, we have shown that measurements of the population of individual magnetic sublevels can be used to take full advantage of the inherent sensitivity to magnetic fields or to the electric dipole moments available from precessing from the $m=0$ state. The larger $F$ is, the greater advantage this type of precession affords compared to Larmor procession. This scheme produces an optimal measurement for $F=1$ but becomes somewhat less sensitive as $F$ increases. We have shown that the total probability to be in even (or odd) magnetic sublevels is insensitive to the tensor Stark shift and the second order Zeeman shift, and that it is possible to combine the measurements of individual sublevels to yield similarly sensitive, purely sinusoidal signals. In general, sensitivities are best when atoms are placed in a superposition of stretched states, and we discuss using such a state for an EDM measurement. When unwanted transverse magnetic fields are too large, it is hard to prepare the superposition of stretched states, but it remains possible to make almost as precise a measurement using precession from the $m=0$ state.

This work was supported by the National Science Foundation (NSF PHY-1607517).

\appendix*
\section{Proof of insensitivity of the probability to be in even magnetic sublevels to quadratic energy shifts in the transverse direction}\label{app:even}
We define $m_e$ and $m_o$ for the even and odd magnetic quantum number $m$ respectively for integer $F$; or $m_e$ and $m_o$ for the even+1/2 and odd+1/2 magnetic quantum number $m$ respectively for half integer $F$. The completeness relation reads:
\begin{equation}\label{completeness}
1=\sum_m\ket{m}\bra{m}=\sum_{m_e}\ket{m_e}\bra{m_e}
+\sum_{m_o}\ket{m_o}\bra{m_o}
\end{equation}
The operator for a measurement of probability to be in the even magnetic sublevels in the $\vect{x}$ basis is $P_{ex}=\sum_{m_e}\ket{m_e}_x\bra{m_e}_x$. Since the Hamiltonian is more conveniently written in the $\vect{z}$ basis, we also write $P_{ex}$ in the $\vect{z}$ basis using the passive rotation $\ket{m}_x=\D_y(\pi/2)\ket{m}_z$, where $\D_y(\theta)$ is the Wigner rotation around $\vect{y}$:
\begin{equation}\label{P in z basis}
P_{ex}=\sum_{m_e}\D_y(\pi/2)\ket{m_e}_z\bra{m_e}_z\D^{-1}_y(\pi/2).
\end{equation}
Because we will work exclusively in the $\vect{z}$ basis henceforth, we will drop the $z$ subscripts for the kets and bras. We will rewrite Eq.~(\ref{P in z basis}) using the following equation:
\begin{align}\label{D and D}
\sum_{m_e}\ket{m_e}\bra{m_e}\D^{-1}_y(\pi/2) = &\frac{1}{2}\sum_{m_e}\{\D_y(\pi/2)+\D^{-1}_y(\pi/2)\}\ket{m_e}\bra{m_e} \nonumber\\
+&\frac{1}{2}\sum_{m_o}\{\D^{-1}_y(\pi/2)-\D_y(\pi/2)\}\ket{m_o}\bra{m_o}.
\end{align}

To prove Eq.~(\ref{D and D}), we expand the rotation operators $\D$ using the completeness relation (\ref{completeness}) and write $\bra{m'}\D_y(\pi/2)\ket{m}$ as $d_{m'm}(\pi/2)$ and $\bra{m'}\D^{-1}_y(\pi/2)\ket{m}$ as $d_{m'm}^\dag(\pi/2)$. $d_{m'm}^\dag(\pi/2)$ is the same as $d_{mm'}(\pi/2)$ because $\D_y$ is unitary and real. The sum over $m_e$ in the right hand side of Eq.~(\ref{D and D}) becomes:
\begin{align}
\frac{1}{2}\sum_{m_e}\{\D_y(\pi/2)+\D^{-1}_y(\pi/2)\}\ket{m_e}\bra{m_e}&=\frac{1}{2}\sum_{m'm_e}\{ d_{m'm_e}(\pi/2)+d_{m_em'}(\pi/2)\}\ket{m'}\bra{m_e} \label{m}\\
&=\sum_{m_e'm_e} d_{m_e'm_e}(\pi/2)\ket{m_e'}\bra{m_e}.\label{separate}
\end{align}
From \ref{m} to \ref{separate}, we separate $m'$ into $m_e'$ and $m_o'$: $\sum_{m'}\ket{m'}\bra{m_e}=\sum_{m_e'} \ket{m_e'}\bra{m_e}+\sum_{m_o'}\ket{m_o'}\bra{m_e}$. The $\ket{m_e'}\bra{m_e}$ terms double and the $\ket{m_o'}\bra{m_e}$ terms drop out because of $d_{mm'}(\pi/2)=(-1)^{m'-m}d_{m'm}(\pi/2)$, which can be shown using the explicit Wigner's formula found in many standard quantum mechanics textbooks such as \cite{Sakurai}. Similarly, the sum over $m_o$ in the right hand side of Eq.~(\ref{D and D}) can be written as:
\begin{align}
\frac{1}{2}\sum_{m_o}\{\D^{-1}_y(\pi/2)-\D_y(\pi/2)\}\ket{m_o}\bra{m_o}&=\sum_{m_e'm_o} d^\dag_{m_e'm_o}(\pi/2)\ket{m_e'}\bra{m_o}.\label{sum over odd}
\end{align}
The sum of Eq. (\ref{separate}) and Eq. (\ref{sum over odd}) yields the left hand side of Eq.~(\ref{D and D}).

Using Eq.~(\ref{D and D}), Eq.~(\ref{P in z basis}) becomes:
\begin{equation}
P_{ex}=\frac{1}{2}+\frac{1}{2}\D_y^2(\pi/2)(\sum_{m_e}\ket{m_e}\bra{m_e}-\sum_{m_o}\ket{m_o}\bra{m_o}).
\end{equation}
$\D_y^2(\pi/2)$ is the same as $\D_y(\pi)$, which can be written explicitly for both integer and half integer $F$ as:
\begin{equation}\label{D pi}
\D_y(\pi)=\sum_m(-1)^{F-m}\ket{-m}\bra{m}.
\end{equation}
It is noted that $\D_y(\pi)$ contains only anti-diagonal components. Using Eq.~(\ref{D pi}), $P_{ex}$ can be written explicitly in the $\vect{z}$ basis as:
\begin{equation} 
P_{ex}=\frac{1}{2}+\frac{1}{2}(-1)^{\lfloor F\rfloor}\sum_m\ket{-m}\bra{m},
\end{equation}
where $\lfloor F\rfloor$ is the floor of $F$. $\lfloor F\rfloor=F$ for integer $F$ and $\lfloor F\rfloor=F-1/2$ for half integer $F$.

Suppose the Hamiltonian can be expanded in powers of $F_z$ (or $m$) in the $\vect{z}$ basis:
\begin{equation}
H=\omega F_z+\alpha_2 F_z^2+\alpha_3 F_z^3+...,
\end{equation}
where $\alpha_n$ are the coefficients of the $n$\textsuperscript{th} order interactions. The measurement operator in the Heisenberg picture is:
\begin{equation}
P_{ex}(t)=e^{iHt/\hbar}P_{ex}e^{-iHt/\hbar}=\frac{1}{2}+\frac{1}{2}(-1)^{\lfloor F\rfloor}\sum_m\ket{-m}\bra{m}e^{-2\omega m-2\alpha_3\hbar^2 m^3...}.
\end{equation}
The diagonal components in $P_{ex}$ do not interact with the Hamiltonian. The anti-diagonal components interact with only the parts of Hamiltonian that are odd with respect to $m$. The parts of the Hamiltonian that are even with respect to $m$, including quadratic interactions, drop out. Therefore, the probability to be in the even magnetic sublevels for an integer angular momentum or in the even+1/2 magnetic sublevels for a half integer angular momentum is insensitive to quadratic energy shifts with respect to the magnetic quantum number $m$.

\end{document}